\documentclass{elsart}

\usepackage{amssymb}
\usepackage{latexsym}
\usepackage{amsmath}
\usepackage{amsfonts}
\usepackage{multirow} 

\usepackage{epsfig}

\begin{document}
\journal{Physica A}
\begin{frontmatter}

\title{Network effects in a human capital based economic growth model}

\author[1]{Teresa Vaz Martins},
\author[2]{Tanya Ara\'ujo},
\author[1]{Maria Augusta Santos}\\
\and \author[2]{Miguel St Aubyn}

\address[1]{Departamento de F\'{\i}sica/Centro de
F\'{\i}sica do Porto,\\ 
Faculdade de Ci\^encias, Universidade do Porto,\\ Rua do Campo
Alegre 687,
4169-007 Porto, Portugal}

\address[2]{ISEG/TULisbon - Technical University of Lisbon, Department of Economics\\
 UECE - Research Unit on Complexity and Economics\\
 R. Miguel Lupi 20,
1249-078 Lisboa, Portugal}


\begin{abstract}
	
We revisit a recently introduced agent model
 [ACS {\bf 11}, 99 (2008)], where economic
 growth is a consequence of education (human capital formation) 
and innovation, 
and investigate the 
influence of the agents' social network, both on an agent's decision to pursue education
 and on 
the output of new 
ideas. Regular and random networks are considered. The results are compared  with the 
predictions of a mean field (representative agent) model.

\end{abstract}


\end{frontmatter}

\section{Introduction}

Recent theoretical macroeconomic growth models emphasize 
the ultimate 
sources of growth. In these so called 'endogenous growth models', 
the long-run 
rate of growth depends on a set of parameters that describe
 some inner
 characteristics of the economy. These characteristics
 may be connected
to preferences (e. g. the relative importance of future
consumption), to some
technological restrictions (e. g. the shape of the aggregate
 production
function), to demographic factors, to the degree of economic
 openness, to
the quality of institutions, and, last but not least, 
to the ability of
producing and adopting innovation, usually related to
education and research
and development activities.

An important fraction of agent-based models in the economics literature concerns 
innovation and diffusion processes (\cite{chiaro} -- \cite{silver2}).  Some of 
these relate to economic growth in an evolutionary economics perspective, as in 
the seminal work of Nelson and Winter \cite{nelson}. The latter emphasized some of the 
neoclassical theory drawbacks concerning the modeling consistency between microeconomic 
technical change and macroeconomic approaches to growth.
The evolutionary approach seeks then a successful combination of the micro 
and aggregate pictures, usually resorting to computer simulation modeling approaches. 
In what concerns economic growth models, Ara\'ujo and St. Aubyn's \cite{araujo} approach, which we 
develop here, innovates within the evolutionary economics based endogenous growth 
literature. They were pioneer in relying on ideas-based macroeconomic growth and on 
the education decision and outcome interactions and dynamics among agents, and not so 
much on the more usually described local innovation processes that swell through the 
economy. In their endogenous growth model ’ideas’ or ’inventions’ are the main growth
 engine. These ideas are non-rival, in the sense that they can be used by some without
 diminishing the possibility of its use by others. Examples relevant to economic growth 
are easy to find -- the power engine, electricity, or the computer come almost immediately
 to mind.  Ideas are produced by skilled
 labor, as proposed by Jones \cite{jones05}. Ara\'ujo and St. Aubyn's model, however, 
departs from more conventional approaches in
 what concerns the way the skilled fraction of the labor force is determined. 
In their agent-based approach, individuals are subject to interaction or 
neighborhood effects, and 
the decision to become educated depends not only on an economic reasoning, but also on a 
degree of social conditioning. As it will become clear later, the skilled or educated 
share of the population is a key factor in what concerns the economy's rate of growth.

Networks of interacting agents play an important modeling role in
fields as diverse as computer science, biology, ecology, economy and
sociology. An important notion in these networks is the {\it
distance} between two agents. Depending on the circumstances,
distance may be measured by the strength of interaction between the
agents, by their spatial distance or by some other norm expressing
the existence of a link between the agents. Based on this notion, global parameters 
have been constructed to
characterize the connectivity structure of the networks. Two of them
are the {\it clustering coefficient} (CC) and the {\it characteristic
path length} (CPL) or geodesic distance. The clustering coefficient measures the average
probability that two agents, having a common neighbor, are
themselves connected. The characteristic path length is the average
length of the shortest path connecting each pair of agents. These
coefficients are sufficient to distinguish randomly connected
networks from ordered networks and from {\it small-world} networks. In
ordered networks, the agents being connected as in a crystal
lattice, clustering is high and the characteristic path length is
large too. In randomly connected networks, clustering
and path length are low, whereas in {\it small-world} networks \cite{Watts1}, \cite{Watts2}
 clustering may be high while the path
length is kept at a low level. Starting from a regular structure
and applying a random rewiring procedure (to interpolate between
regular and random networks) it has been found \cite{Watts1} that
there is a broad interval of structures over which CPL is almost as
small as in random graphs and yet CC is much greater than expected
in the random case and a {\it small-world} network is obtained. An alternative way 
to generate a {\it small-world} network is by 
adding a small number of random short-cuts to a regular lattice.

A small characteristic path length and very high 
clustering are found in real social networks, together with a non-uniform distribution 
of node connectivities. Other distinctive features of this type of networks are positive 
degree correlations (assortative mixing) and the existence of a community 
structure \cite{newman03}, \cite{sen}. 

In the present paper, we investigate the role of the network of interactions in the model 
introduced by Ara\'ujo and St Aubyn \cite{araujo}. The aim is to clarify to what extent the conclusions 
of \cite{araujo} depend on the simplified topology of agents interactions assumed in that
 work. We have simulated the model on networks with some of the features described above, namely on regular square lattices, classical random (Erd\H os-R\'enyi) graphs as well as {\it small-world} networks. An analytical study of the model, using a mean-field-like approximation, is also presented.

The original model definitions and results are reviewed in the next section. In section 
3 we present the mean field results. Simulations on random networks are detailed in 
section 4, and section 5 is devoted 
to a discussion of the results. Conclusions are presented in the final section.

\section{Ara\'ujo and St Aubyn's model}

In the model economy introduced in \cite{araujo} there is a constant population of $N$ individuals (agents), who live for two periods of time; each individual is a {\it junior} in the first period and becomes a {\it senior} in the second period of his life. There is generations overlap and one further assumes that at any time $t$ there are $N/2$ juniors and $N/2$ seniors. Agents are either {\it skilled} or {\it unskilled}; an unskilled junior is part of the unskilled labor force, whereas a skilled junior is a student who postpones joining the (skilled) labor force until he becomes a (skilled) senior. Unskilled juniors will later turn to unskilled seniors.

Denoting respectively by $J_s(t)$, $J_u(t)$, $S_s(t)$ and $S_u(t)$ the total number of skilled juniors, unskilled juniors, skilled seniors and unskilled seniors, one has

\begin{eqnarray}
J_s(t)+J_u(t)&=&S_s(t)+S_u(t)=N/2\\
S_S(t)&=&J_S(t-1)\\
S_u(t)&=&J_U(t-1)
\end{eqnarray}

and the total number of unskilled workers is

\begin{equation}
U(t)= S_u(t)+ J_u(t).
\end{equation}

Individuals live at fixed positions in space -- which is taken in \cite{araujo} as a one-dimensional regular lattice with periodic boundary conditions.

One may then represent the state of agent $i$ ($i=1,...N$) by a 'microscopic' variable $\sigma_i$ which takes one out of four values; $\sigma_i=0$ (skilled junior), $\sigma_i=1$ (unskilled junior), $\sigma_i=2$ (skilled senior) and $\sigma_i=3$ (unskilled senior).

At fixed time intervals ({\it periods}) the states of all agents are updated simultaneously, according to deterministic rules. Agents in junior states turn into the corresponding senior state ($0 \rightarrow 2$, $1 \rightarrow 3$); an agent in a senior state is replaced by either a skilled or an unskilled junior, depending on the result of a 'decision rule'. The decision of a junior agent to follow education is determined both by imitation of his neighbors (neighborhood effect) and by external information (concerning the relative wages of skilled and unskilled workers). According to model \cite{araujo}, a 'newborn' agent $i$ will become a student if there is a weighted majority of skilled agents in his neighborhood. More precisely $\sigma_i=2,3 \rightarrow 0$ if

\begin{equation}
r_w(t) S_{s(i)}(t) > S_{u(i)}(t)
\end{equation}

where $S_{s(i)}$ ($S_{u(i)}$) is the number of senior skilled (unskilled) neighbors of agent $i$ and $r_w(t)$ is the relative weight of the skilled, given by

\begin{equation}
r_w(t) \equiv \alpha \prime \frac{w_s(t-1)}{w_u(t-1)}.
\end{equation}

In (6) $w_s$ ($w_u$) denotes the skilled (unskilled) workers' salary and $\alpha\prime$ is an external parameter, accounting for a bias towards education and corrections due to discount rate \cite{Comment}.  A larger value of $\alpha\prime$ may represent a positive bias towards education and/or that the future is less valued.
If condition (5) is not fulfilled, agent $i$ becomes an unskilled junior ($\sigma_i=2,3 \rightarrow 1$).

{\bf Production}\\
Skilled (intellectual) and unskilled (manual) workers have distinct roles in the model economy: intellectual workers produce {\it ideas} whereas manual workers use existing ideas to produce final goods. Denoting by $A(t)$ the present stock of ideas, the final production function $Y(t)$ is written as 

\begin{equation}
Y(t)=A(t)\,U(t).
\end{equation}

New ideas are created according to 

\begin{equation}
\Delta A(t) \equiv (A(t)-A(t-1)) =  A(t-1)(\delta S_s(t)+ \gamma D(t))
\end{equation}

with
\[D = \frac{1}{S_s} \sum_{i \neq j} \frac{1}{d_{ij}}\]
$d_{ij}$ denoting the {\it distance} between skilled seniors $i$ and $j$. When $\gamma >0$,
 the second term in (8) yields a reinforcement of the output of ideas if skilled workers are 
close to each other -- $\gamma \,(\ge 0)$ is hence called the {\it team effect parameter}.
 Parameter $\delta$ is related to skilled labor marginal productivity.

The total income $Y$ is distributed as wages and split between unskilled and skilled 
labor forces, $Y(t)=Y_u(t)+Y_s(t)$. The share of each class of workers is assumed to
 be the result of a {\it social pact} \cite{MS} and is given by 

\begin{eqnarray}
Y_s(t) =\Delta A(t)\, U(t)\\
Y_u(t) = A(t-1) \,U(t),
\end{eqnarray}

yielding individual salaries
\begin{eqnarray}
w_s(t)=\frac{Y_s(t)}{S_s(t)}=(A(t)-A(t-1))\frac{U(t)}{S_s(t)}\\
w_u(t)=\frac{Y_u(t)}{U(t)}=A(t-1).
\end{eqnarray}

The current value of the relative wage $w \equiv  \frac{w_s}{w_u}$ is taken into account
 whenever a new agent makes his choice regarding education (eq. 5). Notice that, in
 the absence of team effect ($\gamma =0$), 

\begin{equation}
r_w(t)=\alpha \prime  \delta \,U(t-1)
\end{equation}

-- thus the existence of a large number of unskilled workers is a stimulus to education.

It is easily seen from eqs. (7) - (12) that the dynamics described above has two trivial 
absorbing states: $U=0$ and $S_s=0$. The former case implies a collapse of the economy 
($Y=0$) due to over-education -- no manual workers exist -- and may be avoided by a modification 
of eqs (9) and (10) \cite{araujo}. The outcome is less severe when $S_s=0$, in which case
 one gets a stagnation of the economy ($\Delta Y=\Delta A=0$), a situation referred to as
 the {\it poverty trap}. 
The system's asymptotic state depends on 
parameter values and also on the initial configuration of agents. When it does not collapse to an absorbing state, a finite system may reach a 
nontrivial fixed point, with a constant number of skilled and unskilled workers; more often, 
the asymptotic state is characterized by a stationary number 
of unskilled workers, while the number of skilled ones oscillates between two values
 -- see Figure 1 (a). This is a consequence of the somewhat artificial 
junior to senior updating rule \cite{comment2}. Still, one may argue that the appropriate time 
unit is a {\it generation} (equal to two periods)
and consider a state where observables are constant along generations a 'steady state'.

In \cite{araujo} an initial state is generated, with a pre-assigned ratio of skilled 
to unskilled workers, randomly placed on a ring of size $N$. Fixing the size of the
 neighborhood (eq. 4) and the values of the exogenous parameters, a baseline and several
 scenarios were considered. In each case, equations (5) - (12) were iterated till a
 'stationary 
state' was reached, and the process was repeated for other realizations of the initial state.
A segregation of educated and uneducated 
agents in separate domains was observed, specially when team effects were taken into 
account -- see Figure 2. The team effect also yielded higher economic growth and qualification, usually
 together with lower relative salary levels for skilled workers. The steady state was 
found to depend on the initial number of intellectual workers, a poor concentration of 
educated agents leading to lower growth rates and lower percentage of skilled workers.

\section{Mean field (representative agent) approach}

In the model described above, the evolution of the economy is directly determined by
 the overall numbers of skilled and unskilled workers (and by their relative positions 
if the team effect is considered). The random initial position of the four types of 
agents is the sole source of stochasticity in the model, since for a particular 
distribution of agents in space, the decision (to study or not to study) is predictable.

In mean field approximations, one neglects fluctuations of the microscopic variables and replaces the detailed form of their mutual interactions by the interaction of a single microscopic variable with a uniform 'mean field', which depends on the state of the system.
In the present case, neglecting the differences in the local neighborhood would lead to one of the two trivial outcomes, 'all educated' or 'all uneducated'. One can, however, obtain a related non trivial model, within the mean field spirit, by considering a single {\it representative agent} who takes the decision of whether or not to follow education stochastically, with relative probabilities depending on the total number of skilled to unskilled workers at that time. In this {\it mean field model} (MF)
 one has uniform initial conditions (or neighborhood) and probabilistic decision making, instead of the random initial conditions and deterministic decision of the original model.
More specifically, we set

\begin{equation}
\frac{J_s(t)}{J_u(t)}=\frac{p_s(t)}{1-p_s(t)},
\end{equation}
where $p_s$ is the probability that a junior decides to study, and assume the decision probabilities depend on the weighted ratio of skilled to unskilled seniors

\begin{equation}
\frac{p_s(t)}{1-p_s(t)}=r_w(t)\, \frac{S_s(t)}{S_u(t)}= r_w(t) \,\frac{J_s(t-1)}{J_u(t-1)}
\end{equation}

($t \ge 1$; $r_w$ is the relative weight given to the skilled, as before). Substituting eq. (15) into eq. (14) one obtains a deterministic nonlinear equation \cite{commentMF} with a stable nontrivial solution, as shown below.

A recursion relation is easily written for $R(t)\equiv \frac{J_s(t)}{J_u(t)}$:
\begin{equation}
R(t)= \lambda \,\,\frac{N}{2}\, (\frac{1}{1+R(t-1)}+
\frac{1}{1+R(t-2)})\,\,R(t-1)
\end{equation}
where $\lambda$ is the product of the education bias by the efficency of skilled workers ($\lambda \equiv \delta \alpha\prime$).
Provided $J_u(0) \neq 0$, this equation has stationary solutions $R^*=0$ or
\begin{equation}
R^*=\lambda N -1,
\end{equation}

independent of initial conditions. From equations (14) - (17) one easily gets the 
following (MF) stationary values:\\

relative wages 
\begin{equation} 
w^*= 1 / \alpha \prime 
\end{equation}

number of unskilled workers
\begin{equation} 
U^*= 1/ \lambda
\end{equation}

number of skilled workers 
\begin{equation}
S_s^*=\frac{N}{2}-\frac{1}{2\lambda}
\end{equation}

growth rate 
\begin{equation}
\Delta Y/Y= \delta\, S_s^*.
\end{equation}

The requirement $U^* \leq N$ puts a limit on parameter values: $\lambda \ge 1/N$. The fact
 that the stationary number of unskilled workers is independent of $N$ (eq. (19)) is obviously
 a spurious result of the mean field approximation! The prediction for the other observables
 is, however, not far from the simulation results for a regular lattice -- see Table 2 -- 
suggesting that, in the absence of team effect, the model introduced in \cite{araujo} 
is well approximated by the mean field model, exception made of the steady state growth 
rate dependency on initial values.

\section{Simulations on complex networks}

The ring topology used in \cite{araujo} is obviously a simplified description of the structure of the social interactions between agents. The mean field approach, on the other hand, assumes that the underlying network is a complete graph where each individual is influenced by all the others -- which is only a good approximation for well-mixed populations. In this section we present the results of model simulations on more realistic social networks.

As described in section 2, agents live on the nodes of a network, which we call the {\it influence network}, since a 'newborn' agent is influenced by his neighborhood on this network, in his decision concerning education.
When team effects are considered (see eq. (8)), the output of ideas is enhanced if the
 'distance' between pairs of skilled workers is small. This distance may be the euclidean
 one on a regular lattice or the shortest path along links of a random network. One may 
even consider a {\it collaboration network} among senior skilled agents, distinct from
 the underlying {\it influence network}. Take for example a situation where education 
is decided on a family/local neighbors basis, whereas intellectual workers collaborate 
with colleagues from a distant town by e-mailing or indirectly through mutual acquaintances.
 This case can be modeled by taking a square lattice (with a neighborhood of size $z=4$ or $z=8$) as influence network, plus a small-world for the collaboration network.

{\bf Regular lattices}\\
A comparison was made between the results for a ring and a square lattice with periodic
 boundary conditions and the same number of neighbors ($z=4$ or $z=8$).

{\bf Random networks}
\begin{itemize}

\item {\bf classical random graph (CRG)}\\
In each run, a connected graph was generated with $N$ vertices and $L=Nz/2$ edges through
 a {\it random graph process}: a randomly chosen pair of vertices is linked by an edge 
and the process is repeated till the pre-defined number of edges, $L$, is obtained 
(disregarding graphs with self-loops or multiple edges).

\item {\bf small-world networks (SW)}\\
Taking a ring or a square lattice as 'mother lattice', short-cuts between randomly chosen 
nodes were added with (small) probability $P$ per regular link of the underlying lattice 
($L \prime =LP$ short-cuts). We have considered the following cases: i) only the {\it influence
 network} is SW (the additional random links are only relevant for the education decision),
 ii) only the {\it collaboration network} is a SW (only senior skilled agents may get additional
 random links) 
and iii) both networks are SW.
The addition of random links is repeated every half generation, so a senior agent keeps the
 regular contacts he had while a junior but is free to establish new short-cuts, and a newborn
 junior does not inherit random links.

\end{itemize}

In our simulations we have used $N=400$ agents and chosen parameter values which yield 
{\it reasonable} stationary growth rates (we verified that a change in parameter values
 does not produce 
 qualitatively distinct conclusions).
A calibration was done using mean-field equations (17) - (21), assuming a $3\%$ annual growth rate 
with a half generation of $25$ years (or $\frac{\Delta {commentMF}Y}{Y}=1.0938$) and fixing $U^*=N/2$.
 One obtains the {\bf baseline} parameter values: $\delta=0.011$, $\alpha \prime =0.45$ 
($\lambda=0.005$), $\gamma=0$. To study team effects (scenario 5 in \cite{araujo})
 we took $\gamma=0.05$.

Simulations started with $U(0)=N/2$ unskilled workers randomly placed on the network
 nodes. Averages
 over up to $1000$ samples were taken for each set of parameters (neglecting samples which
 collapsed to an 
absorbing state). 
Typical evolutions of the populations of skilled and unskilled workers in single 
runs are represented in Figure 1, together with the respective average over runs, for the case of
(a) a ring and (b) a small-world network (case ii), with team effect parameters. 
The period-two 
oscillatory behaviour
of the number of skilled seniors, discussed in section 2, yiels synchronous oscillations of
the product growth rate, but with rather small amplitude (roughly equal to $\delta=0.011$ times 
the skilled workers' amplitude). 
An illustration of an initial and final network configurations, where domain formation is evident, 
is presented in Figure 3, for the case of a small-world built upon a ring. 
For the parameter values used in the simulations, relaxation times varied from about $10$ to $300$ 
periods (the relaxation is slower when the {\it influence network} is a small-world with
 few short-cuts). The overall picture is, however, qualitatively similar for all types of networks, with
 or without team effect. Simulation results for the 'steady 
state' values of observables are summarized in Tables $1$ and $2$.

\begin{figure}
\begin{center}
\includegraphics[width=6.0cm]{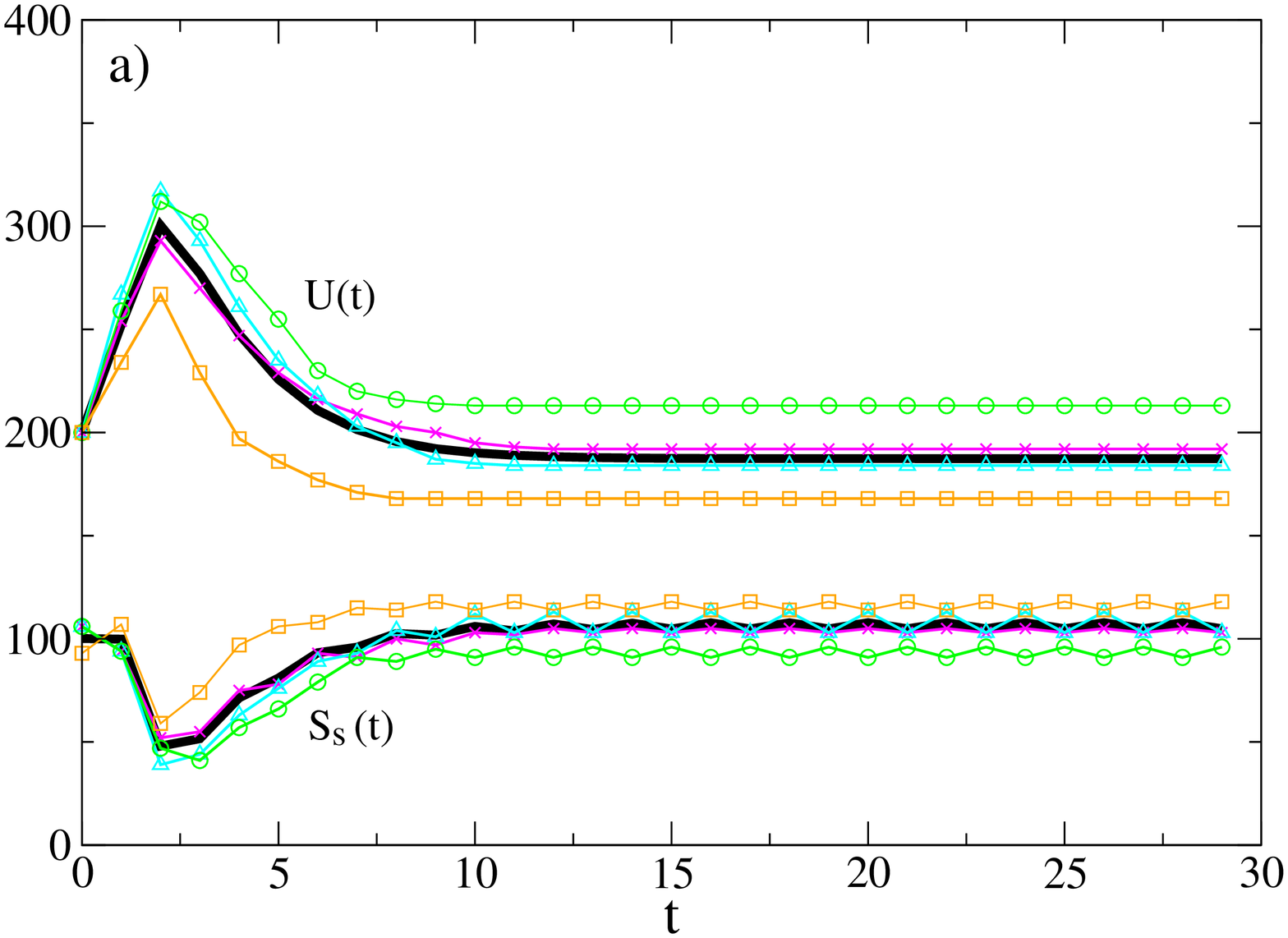}
\hspace{1cm} 
\includegraphics[width=6.0cm]{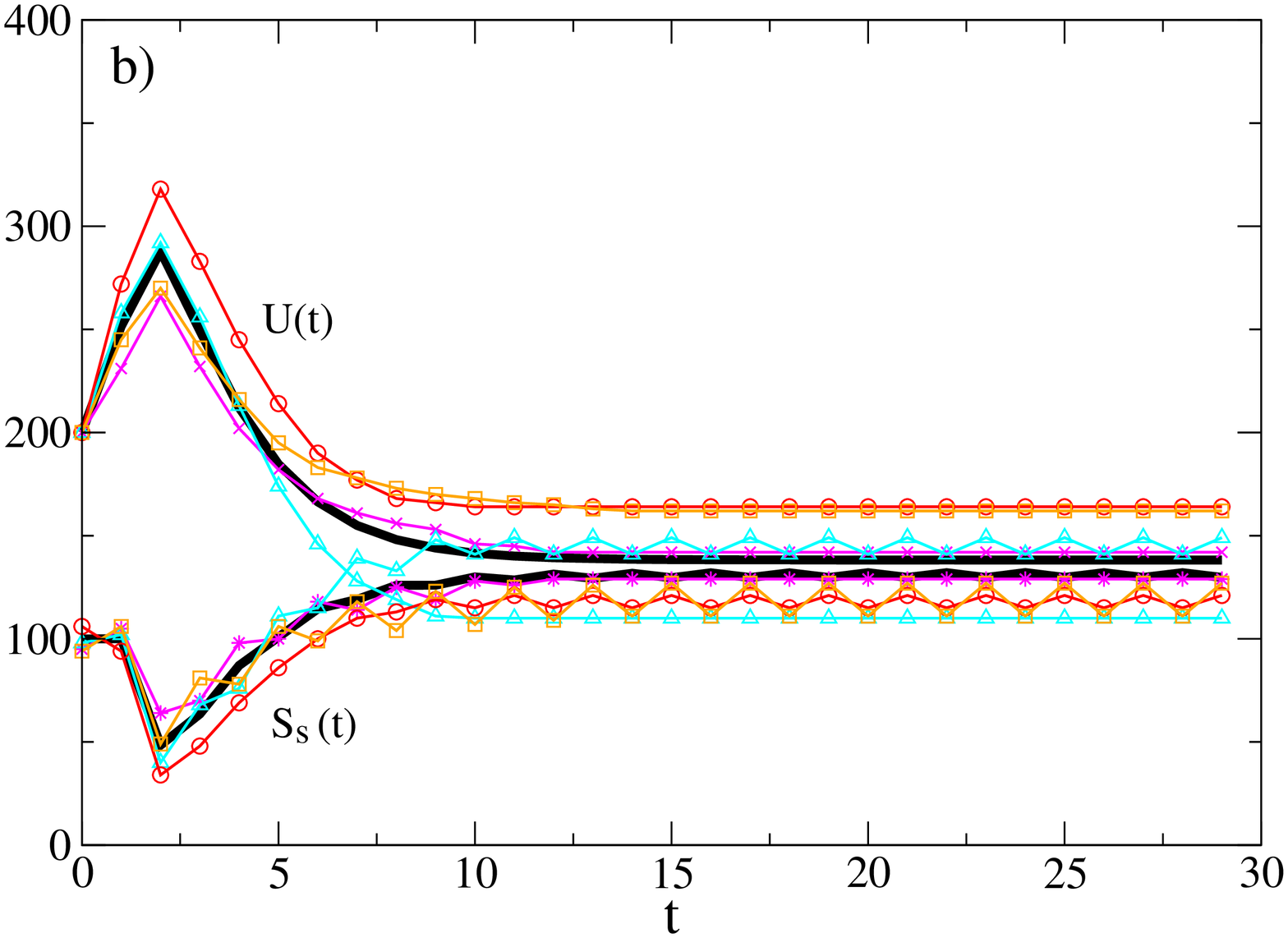}
\caption{Evolution of the number of unskilled workers $U(t)$ (upper curves) 
and skilled workers $S_s(t)$ (lower curves) for (a) a ring with $z=6$ neighbours, and (b) a small-world
 built on 
the ring with random links added with probability $P=0.03$ case ii); team effect parameters (see text).
Symbols (and colours online) represent distinct 
runs; the full (black) lines are averages over $1000$ runs.}
\label{fig1}
\end{center}
\end{figure}
\vspace{1cm}

\begin{figure}
\begin{center}
\includegraphics[width=6.6cm]{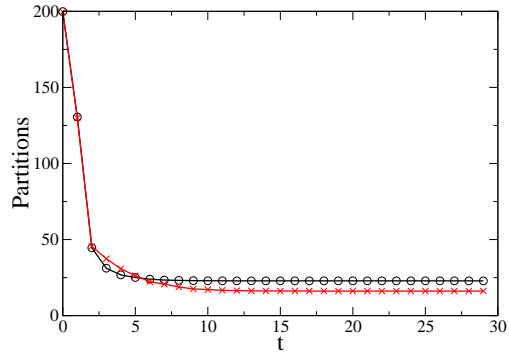}
\caption{Average number of walls (partitions) between domains of skilled/unskilled agents as a 
function of time. Ring with $z=6$ neighbours, baseline parameters (circles) and with team effect (crosses).}
\label{fig2}
\end{center}
\end{figure}
\vspace{1cm}

\begin{figure}
\begin{center}
\includegraphics[width=5.cm]{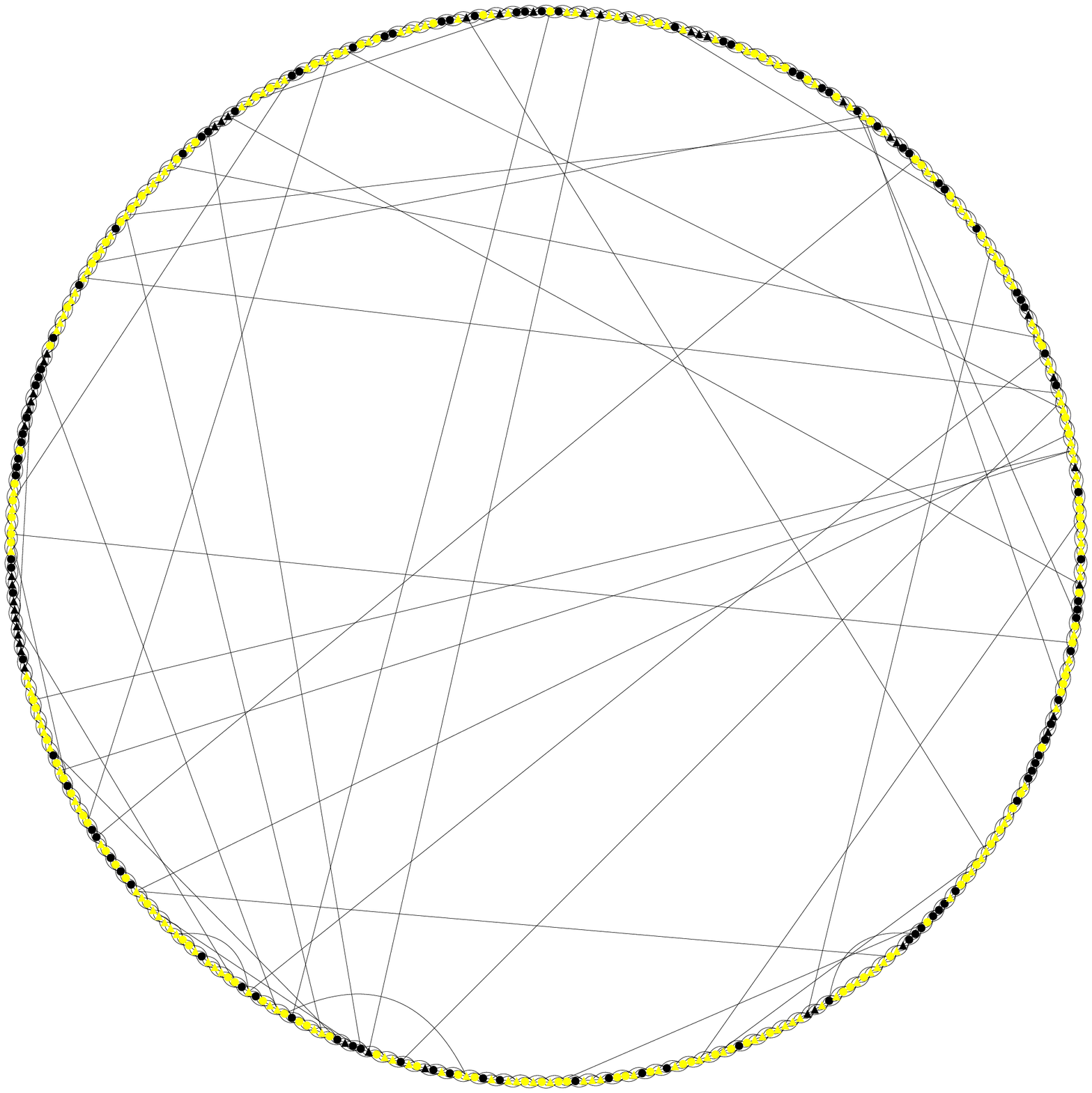}
\hspace{1cm} 
\includegraphics[width=5.cm]{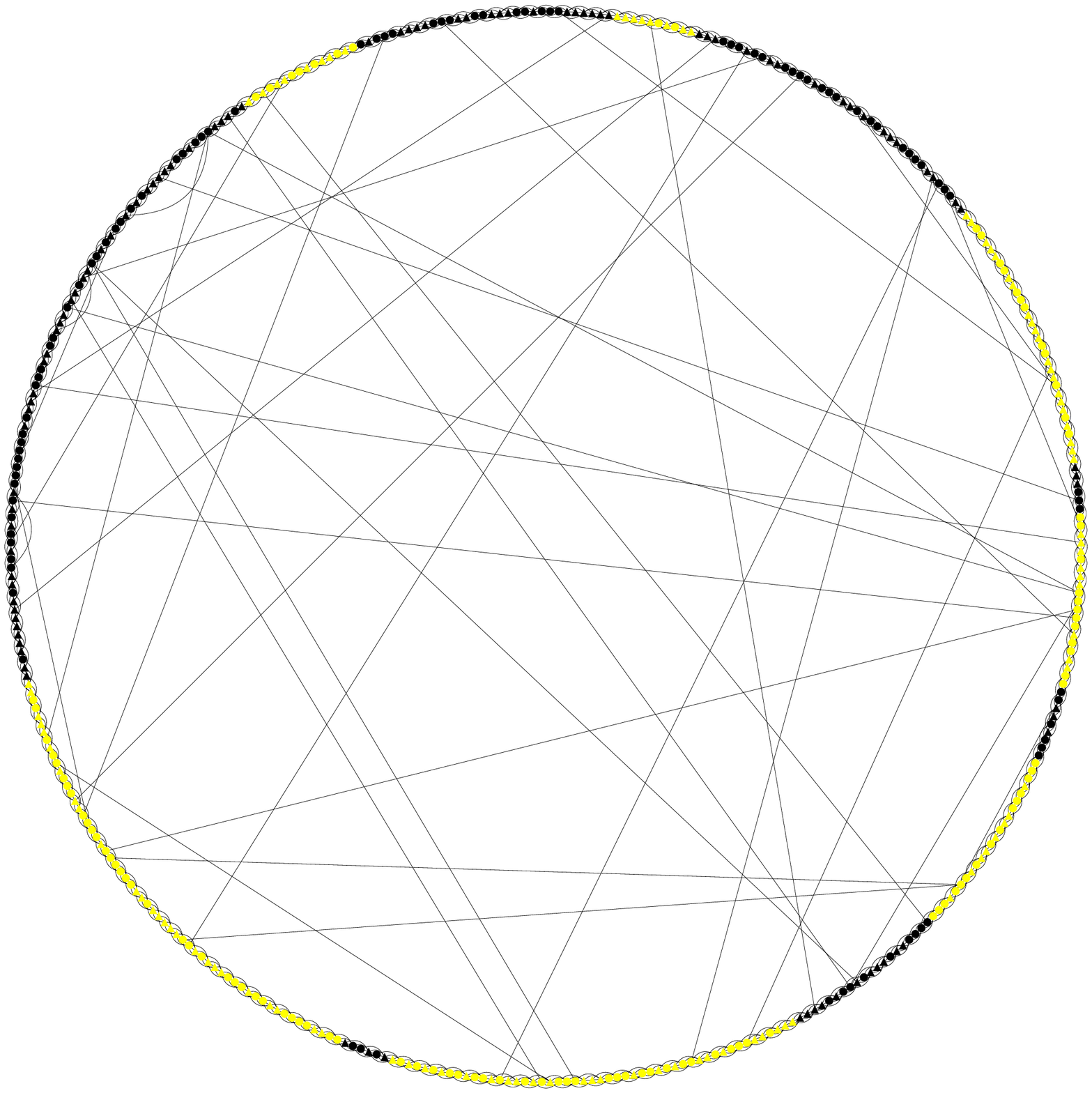}
\caption{Initial (left) and final (right) network configurations -- SW case i), random links added with probability $P=0.03$.
Nodes represent skilled (black) or unskilled (yellow)
 agents on a ring with $z=6$ regular links per node (hardly visible) plus some short-cuts; 
baseline parameters (see text).}
\label{fig3}
\end{center}
\end{figure}

\newpage
\section{Discussion}
Our simulations confirm the statement \cite{araujo} that, within this model, in 
the long run, higher growth rates
 are associated with higher qualification (more skilled workers and less unskilled ones) and 
stronger segregation by educational level. Usually this also implies smaller (relative) 
salaries for the intellectual workers.

From Tables 1 and 2, one sees that {\bf baseline} results are rather similar for a ring or a square
 lattice with the same number of neighbors and they also do not differ significantly for a 
random graph with the same average degree. In contrast, a clear increase in qualification
 is obtained if the influence network is a small-world (specially for higher $P$ values) 
\cite{comment}.

When {\bf team effects} are taken into account, the effect of random inter-agent connections is evident
 from the much higher growth rates and larger fraction of skilled workers. Since collaboration 
networks are often found to be scale-free \cite{newman01}, we also ran simulations where the 
number of links between senior intellectuals had a power-law distribution (with exponent $1$).
 Those were not true scale-free networks, since the number of nodes was just about
 $160$, but the purpose was to check if significant changes were seen. The results were  very 
similar to the SW-case iii) with $P=0.1$, except for a larger growth rate ($\Delta Y/Y=5.85$).
In the case of regular lattices, team effects are much stronger in 2d than in 1d (due to the higher
 clustering of the square lattice), yielding increased growth/qualification in two dimensions.
The tendency for segregation is stronger on the (more realistic) 2d substract, as shown
 by the smaller number of partitions (domains) - see Table 2.

We have run simulations for higher numbers of agents and obtained qualitatively the same
behavior, but the fraction of skilled workers increases with $N$, leading to higher 
growth rates. Theoretical endogenous growth models often display this type of behavior,
 with larger economies tending to have higher growth rates \cite{jones99}. Still, $N=400$ 
is too small 
for most realistic economies, so the values presented in Tables $1$ and $2$ should be 
taken just 
as indicative for the sake of comparison, and not on absolute terms.

\section{Conclusion}

In this work we have studied a simple agent model \cite{araujo} designed to explain how
 economic growth
can be the result of innovation introduced by a fraction of the population with higher
education. The original unidimensional model was generalized by assuming that agents 
interactions took place on a more realistic type of social network.

The effect of assuming a heterogeneous interaction network may be summarized, in
 broad terms, 
as leading to a steady state with higher growth and qualification and a stronger
segregation of skilled and unskilled individuals. As expected, this 
is specially notorious when team effects are considered, due to the shorter average 
distance between agents.

We have also derived and solved a related mean field model, which is able to
reproduce some important results of the original model (without team effect). Note,
 however, that the mean field approximation leaves out the steady state growth rate
 dependency on initial values. Also, by
 assuming heterogeneous agents, one is
able to reproduce cluster
 formation and to take into account a team contribution.

It will be interesting to consider a more microscopic version of the model, with
 heterogeneous salaries
 depending on the agent's productivity (as well as on overall economic state).
 The artificial
 decision rule (5), may be replaced by the agent making his choice according to the sign of 
a {\it local field}, as usually done in models of social influence with binary
 decision \cite{rfim}.
The question is then how to relate agent variables to production and innovation in an 
economically sensible way.

{\underline{Acknowledgment}}
T. A. and M. S. acknowledge support from FCT (Portugal),
Project PDCT/EGE/60193/2004. T.V.M. is supported by FCT (Portugal) through 
Grant No.SFRH/BD/23709/2005.\\ 
UECE and Centro de F\'{\i}sica do Porto are
supported by FCT (Funda\c c\~ao para a Ci\^encia e a Tecnologia, Portugal),
financed by ERDF and Portuguese funds.

\newpage
\begin{table}
\caption{'Steady state' results for a ring, a small world network (SW) built on the ring  
(cases i)- iii) in the text), a random graph (CRG) and mean field (MF). The number of domains of 
skilled/unskilled along the ring is shown in 'Partitions'; 'Team Eff' is the value 
of $\gamma D$ in eq.(8).
N=400, $z=6$, $\alpha'=0.45$, $\delta=0.011$, $\gamma=0.05$ for team effect.}
\label{tab:1}       
\begin{minipage}{\textwidth}
\begin{tiny}
\vspace{0.5cm}

\begin{tabular}{|c|c|c|c|c|c|c|c|c|c|c|c|c|c|}
\hline

 \multirow{3}{*}{} & \multicolumn{2}{c|}{Ring} & \multicolumn{4}{c|}{SW - Decision} & \multicolumn{2}{c|}{SW - Team Effect} & \multicolumn{2}{c|}{SW - Dec.+ Team} & \multicolumn{2}{c|}{CRG} & MF \\\cline{2-14}

 & \multirow{2}{*}{Base} & \multirow{2}{*}{Team}  & \multicolumn{2}{c|}{p=0.03} & \multicolumn{2}{c|}{p=0.1} & \multirow{2}{*}{p=0.03} & \multirow{2}{*}{p=0.1} & \multirow{2}{*}{p=0.03} & \multirow{2}{*}{p=0.1} & \multirow{2}{*}{Base} & \multirow{2}{*}{Team} &\multirow{2}{*}{Base} \\\cline{4-7}

 & &   & Base & Team & Base & Team &  & &  & & & &\\
\hline

$\Delta Y/Y$ & 0.666 & 1.79 & 1.06 & 2.23 & 1.09 & 2.25 & 2.78 & 3.33 & 3.16 & 3.78 & 0.67 & 3.67 & 1.09 \\ \hline

$w$ & 3.05 & 3.20 & 2.26 & 2.24 & 2.22 & 2.21 & 3.11 & 3.09 & 2.20 & 2.17 & 3.05 & 3.11 & 2.22 \\ \hline

$S_s$ & 60.6 & 104.9 & 96.8 & 132.9 & 99.1 & 133.1 & 128.9 & 136.1 & 148.3 & 155.1 & 61.1 & 140.2 & 100 \\ \hline

$U$ & 277.5 & 187.3 & 205.5 & 133.6 & 202.1 & 134.9 & 141.0 & 126.3 & 102.4 & 89.3 & 277.4 & 119.1 & 200 \\ \hline

Partitions & 23.1 & 16.4 & 14.7 & 12.4 & 16.7 & 15.2 & 14.3 & 13.3 & 11.1 & 12.1 & 22.1 & 12.7 & - \\ \hline

Team Eff & - & 0.635 & - & 0.770 & - & 0.765 & 1.36 & 1.83 & 1.56 & 2.06 & - & 2.12 & - \\ \hline

\end{tabular}
\end{tiny}
\end{minipage}
\end{table}

\begin{table}
\caption{Comparison of 'steady state' results on $1d$ and $2d$ substracts and mean field (MF). 'Partitions' is the number of clusters of skilled/unskilled on the square lattice; 
'Team Eff' is the value 
of $\gamma D$ in eq.(8).
$N=400$, $z=4$, $\alpha'=0.45$, $\delta=0.011$, $\gamma=0.05$ for team effect.}
\label{tab:2}       

\begin{minipage}{\textwidth}
\begin{tiny}
\vspace{0.5cm}

\begin{tabular}{|c|c|c|c|c|c|c|c|c|c|}
\hline

 \multirow{2}{*}{} & \multicolumn{2}{c|}{Ring} & \multicolumn{2}{c|}{Ring/SW -Team Effect} &  \multicolumn{2}{c|}{Square} & \multicolumn{2}{c|}{Square/SW -Team Effect} & MF \\\cline{2-10}

& Base & Team & p=0.03 & p=0.1 & Base & Team & p=0.03 & p=0.1 & Base \\ \hline

$\Delta Y/Y$ & 0.760 & 1.25 & 2.04 & 2.61 & 0.762 & 1.92 & 2.37 & 2.75 & 1.09  \\ \hline

$w$ & 2.79 & 3.35 & 3.30 & 3.28 & 2.82 & 3.24 & 3.01 & 2.97 & 2.22 \\ \hline

$S_s$ & 69.1 & 81.1 & 108.6 & 120.5 & 69.3 & 107.4 & 120.7 & 128.2 & 100 \\ \hline

$U$ & 253.6 & 230.6 & 175.7 & 151.1 & 256.2 & 180.9 & 153.1 & 138.2 & 200 \\ \hline

Partitions & 35.0 & 31.9 & 25.9 & 24.3 & 9.3 & 6.7 & 7.4 & 7.9 & - \\ \hline

Team Eff & - & 0.357 & 0.845 & 1.29 & - & 0.742 & 1.04 & 1.34 & - \\ \hline

\end{tabular}
\end{tiny}
\end{minipage}
\end{table}


\begin{thebibliography}{10}
\bibitem{Comment} In the notation of \cite{araujo} $\alpha \prime  \equiv \alpha / \beta(\rho)$
 where $\beta(\rho)$ is an increasing function of the discount rate $\rho$.
\bibitem{MS} The social pact ensures that unskilled workers earn what would have been 
produced if there were no skilled workers in society, and that skilled workers get
 the extra income derived from new ideas.
\bibitem{comment} Note that the SW built by adding shortcuts to a regular lattice with $z$
 neighbors, has slightly higher average degree $<z>=z (1+P/2)$.
\end{thebibliography}
\end{document}